\documentclass[%
reprint,
twocolumn,
showpacs,
amsmath,amssymb,amsfonts, 
aps,
prl,
floatfix,notitlepage,latexsym%,longbibliography
]{revtex4-1}

\usepackage{amsmath,amssymb,amsthm,bm,braket,ascmac,bbm,color}
\usepackage{graphicx}
\usepackage{comment,xcolor}

\usepackage{dcolumn}% Align table columns on decimal point
\usepackage{bm}% bold mathematica
\usepackage{braket}
\usepackage[dvipdfmx]{attachfile2}
\usepackage{subfigure}
\usepackage{comment}

\newcommand{\eV}{\,\mathrm{eV}}
\newcommand{\mvcm}{\,\mathrm{MV/cm}}

\newcommand{\up}{\mathrm{up}}
\newcommand{\low}{\mathrm{low}}
\newcommand{\htot}{\hat{H}_\text{total}}
\newcommand{\Jpara}{J_x}
\newcommand{\Jperp}{J_y}
\newcommand{\AHH}{A^\text{HH}}
\newcommand{\SHH}{S^\text{HH}}

\newcommand{\dd}{\mathrm{d}}
\newcommand{\ii}{\mathrm{i}}

\newcommand{\bk}{\bm{k}}
\newcommand{\mR}{\mathcal{R}}

\newcommand{\hl}[1]{\textcolor{black}{#1}}

\begin{document}
\title{
High-order nonlinear optical response of a twisted bilayer graphene}
\author{Tatsuhiko N. Ikeda}
\affiliation{Institute for Solid State Physics, University of Tokyo, Kashiwa, Chiba 277-8581, Japan}
\date{\today}
\begin{abstract}
Focusing on the twist angle for the minimal commensurate structure,
we perform nonperturbative calculations of electron dynamics in the twisted bilayer graphene (TBG) under intense laser fields. We show that the TBG exhibits enriched high-harmonic generation that cannot occur in monolayer or conventional bilayers. We elucidate the mechanism of these nonlinear responses by analyzing dynamical symmetries, momentum-resolved dynamics, and roles of interlayer coupling. Our results imply nonlinear ``Opto-twistronics'', or controlling optical properties of layered materials by artificial twists.
\end{abstract}
\maketitle

\begin{figure}[t]
\center
\includegraphics[width=\columnwidth]{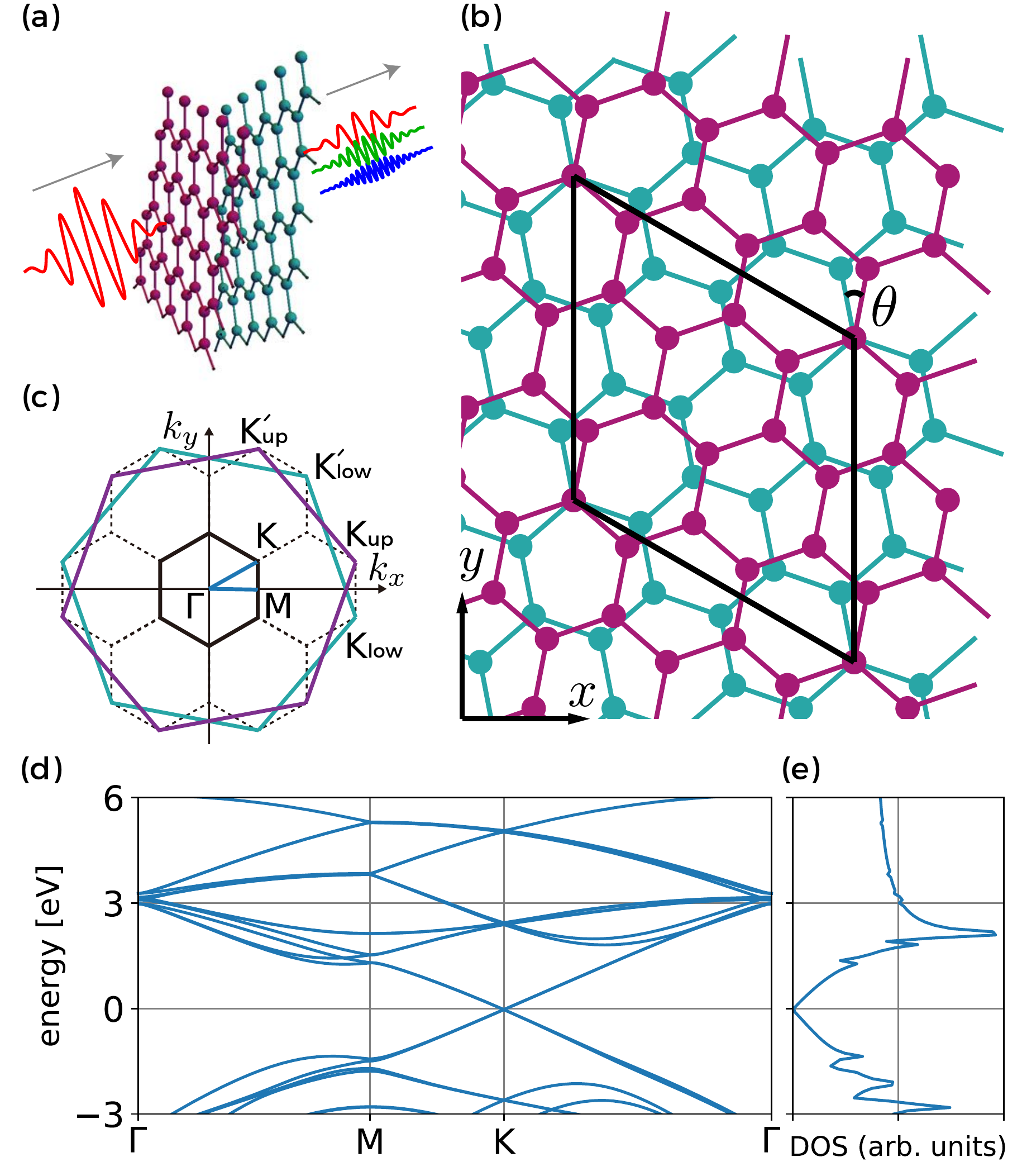}
\caption{
(a)
Schematic illustration of high-harmonic generation
in twisted bilayer graphene.
(b) Top view of the lattice structure of our TBG.
The upper and lower layers rotate respectively by the angles $-\theta/2$ and $\theta/2$
with $\theta=21.79^\circ$ around a common A site.
The parallelogram shows the unit cell involving 28 sites.
(c) The central solid hexagon shows the first Brillouin zone (BZ)
for the superlattice,
and the dotted ones the other BZs. The larger hexagons represent the BZs
for the upper and lower graphenes.
(d) Electronic band structure around the Fermi energy (set to zero)
together with (e) the corresponding density of states~\cite{Li2010,Moon2012}.
}
\label{fig:band}
\end{figure}

{\it Introduction.}---
Nonlinear optical response of materials~\cite{Franken1961,Boydbook}
in intense optical fields
have attracted growing attention
since the invention of laser~\cite{Schawlow1958,Maiman1960}.
The high-harmonic generation in solids~\cite{Ghimire2011,Schubert2014}
is the prototypical nonlinear
phenomenon
(see Fig.~\ref{fig:band}(a)),
and has seen a remarkable development
in the last decade~\cite{Hohenleutner2015,Ndabashimiye2016,You2017,Higuchi2017,Kaneshima2018}.
This phenomenon
has attracted interest
not only for compact frequency converter applications~\cite{Ghimire2014a,Ghimire2019}
but also as a probe of electron
dynamics in intense optical fields~\cite{Vampa2015a}.
Among various systems such as
semiconductors~\cite{Golde2006,Golde2008,Wu2015,Vampa2015d,Ikemachi2017,Osika2017,Du2017,Catoire2018,Ikeda2018b,Navarrete2019,Xia2020}, superconductors~\cite{Matsunaga2014,Kawakami2018,Nakamura2020},
strongly correlated systems~\cite{Ikemachi2018,Murakami2018b,Imai2020,Lysne2020,Roy2020,Wang2020},
quantum magnets~\cite{Lu2017,Takayoshi2019,Ikeda2019}, and topological insulators~\cite{Bauer2018,Jurss2019},
Dirac materials have turned out
to have extremely-large nonlinear susceptibility
from the mid-infrared~\cite{Yoshikawa2017,Baudisch2018,Jiang2018} down to the teraherz~\cite{Hafez2018,Cheng2020,Kovalev2019} frequency regimes.
In particular, nonlinear response of graphene
has been studied extensively~\cite{Mikhailov2007,Wright2009,Ishikawa2010,Al-Naib2014,Rostami2020}.

Very recently, the twisted bilayer graphene (TBG)
has opened a new avenue in physics of Dirac electrons in condensed matter~\cite{Cao2018,Cao2018a}.
The TBG consists of two sheets of graphene vertically stacked
with an artificial twist angle,
which enables us to manipulate
electronic properties of layered materials~\cite{Tran2019}
as sometimes called ``twistronics''~\cite{Carr2017}.
The twist angle brings about physical phenomena
not present in a monolayer graphene
such as superconductivity~\cite{Cao2018,Lu2019,Yankowitz2019,Isobe2018}, Mott-like insulating states~\cite{Cao2018}, to name a few.
Microscopic theories~\cite{Bistritzer2011,Shallcross2010,Moon2012,Koshino2018,Po2018} of the TBG have developed, and many active studies are going on to discover and understand novel phenomena~\cite{Chou2020,Novelli2020}.

However, the nonlinear optical response of the TBG,
or nonlinear ``Opto-twistronics'',
has not yet been explored well.
One theoretical challenge is that
numerous electronic bands are involved in the TBG
due to the large unit cell of the
moir\'{e} structure.
Recently, the Floquet band engineering has been proposed
based on the tight-binding model~\cite{Topp2019}
and the low-energy
effective Hamiltonian involving a few bands~\cite{Katz2019,Vogl2020a,Vogl2020}.
%Yet, this approach can be less accurate for stronger optical fields,
%for which nonlinear optical effects are relevant.
Another approach is the perturbation theory for the optical field.
In this approach, 
the circular photogalvanic effect~\cite{Gao2020,Otteneder2020},
one of the lowest-order nonlinear effects,
has been found, but analyzing higher-order effects would become
more challenging.

In this \hl{Rapid Communication},
by restricting ourselves to a twist angle
resulting in the minimal number of bands,
we show that the TBG exhibits higher-order
nonlinear responses that cannot happen
in monolayer or conventional AA- or AB-stacked bilayers.
The restriction enables the nonperturbative calculation
of electron dynamics in the full number of bands.
We explain the nonlinear responses characteristic to the TBG
by the dynamical symmetries of the Hamiltonian,
\hl{where the key is that the TBG has a smaller point-group symmetry
than monolayer or conventional bilayers.
Thus, the qualitative results shown
in this work also apply to most twist angles leading to the same point-group symmetry.
We also elucidate the mechanism of the nonlinear responses of the TBG}
by the reciprocal-space-resolved analysis
and the decomposition of the electric current
into the intralayer and interlayer contributions.

{\it Model and setup.}---
We begin by defining the lattice structure
of the TBG that we study in this work.
We consider two graphenes, or honeycomb lattices,
on top of each other, i.e., the AA-stacked bilayer.
We let $\bm{r}^{(l)}_i$ denote each site, 
where $l$ ($=\up$ or $\low$) labels each layer
and $i$ does each site within the layer.
Thus $\bm{r}^{(\up)}_i$ and $\bm{r}^{(\low)}_i$
share their $x$ and $y$ components,
but differ in their $z$ components:
$[r^{(\up)}_i]_z-[r^{(\low)}_i]_z=d_0$ ($\forall{i}$)
with $d_0$ being the interlayer distance.

The minimal commensurate TBG is obtained
by rotating the upper (lower) layer by an angle $-\theta/2$ ($\theta/2$)
with $\theta=21.79^\circ$
about the $z$-axis as illustrated in Fig.~\ref{fig:band}(b).
Thus each site of the TBG is located at $\bm{R}^{(\up)}_i=\mathcal{R}_z(-\theta/2)\bm{r}^{(\up)}_i$ and $\bm{R}^{(\low)}_i=\mathcal{R}_z(\theta/2)\bm{r}^{(\low)}_i$, where $\mathcal{R}_z(\varphi)$ represents the $3\times3$ rotation matrix about the $z$-axis
by angle $\varphi$.
Here the commensurability means the presence of
the exact discrete translation symmetry,
and the unit cell contains 28 sites for $\theta=21.79^\circ$.
For other twist angles, the TBG has incommensurate structures
or commensurate ones with larger unit cells.
One exception is the $60^\circ$-twist,
which gives the conventional AB-stacked bilayer.
However, as we will see below, the nonlinear optical responses
for this case are similar to the monolayer or the AA-stacked bilayer.

To describe the quantum states of the electrons on the TBG,
we adopt the tight-binding model of Refs.~\cite{Moon2012,Moon2013}
\begin{align}
    H_\text{TBG}=-\sum_{(i,l),(i',l')}t(\bm{R}^{(l)}_i,\bm{R}^{(l')}_{i'})\ket{\bm{R}^{(l)}_i}\bra{\bm{R}^{(l')}_{i'}}+\text{h.c.},
\end{align}
where $\ket{\bm{R}^{(l)}_i}$ denotes the Wannier state
at position $\bm{R}^{(l)}_i$.
The transfer integral
$t(\bm{R}^{(l)}_i,\bm{R}^{(l')}_{i'})$
between $\bm{R}^{(l)}_i$ and $\bm{R}^{(l')}_{i'}$
depends only on the distance $|\bm{R}^{(l)}_i-\bm{R}^{(l')}_{i'}|$
and its parametrization is taken from Refs.~\cite{Moon2012,Moon2013}.
By the Fourier transform in the $xy$-plane,
we obtain the reciprocal-lattice representation:
$H_\text{TBG}=\sum_{\bm{k},\mu,l,\nu,l'} h_{\mu l,\nu l'}(\bm{k})\ket{\bm{k};\mu,l}\bra{\bm{k};\nu,l'}$,
where $\bm{k}=(k_x,k_y)$ is the two-dimensional wave vector
and the pair $(\mu,l)$ ($\mu=1,2,\dots,14$ and $l=$ up or low)
serves as the internal degree of freedom
corresponding to each site in the TBG unit cell.

The band structure of our TBG is obtained from the eigenvalues of the $28\times28$ Hamiltonian matrix $h_{\mu l,\nu l'}(\bm{k})$ and shown in Fig.~\ref{fig:band}(d) (see also Ref.~\cite{Moon2012}). Throughout this work, we assume the half-filling and set $E_F=0$. We remark that the Dirac cone at the $K$ point is approximately doubly degenerate
besides the spin degeneracy. This degeneracy comes from the Dirac electrons of the upper and lower layers. The interlayer coupling does not affect these Dirac electrons much but causes band splittings away from the $K$ point.

Now we introduce the coupling of the TBG
to the laser propagating in the $z$-direction.
Considering that the laser wavelength is larger enough
than the interatomic distances,
we assume that the 
laser electric field $\bm{E}(t)=(E_x(t),E_y(t),0)$
is homogeneous.
Then the coupling energy is given by
\begin{align}
    H_\text{ext}(t) = \sum_{(i,l)}e\bm{E}(t)\cdot\bm{R}^{(l)}_i\ket{\bm{R}^{(l)}_i}\bra{\bm{R}^{(l)}_{i}},
\end{align}
where $e$ is the elementary charge.
The total Hamiltonian
in the Fourier representation is given by
$\htot(t)\equiv H_\text{TBG}+H_\text{ext}(t)=\sum_{\bm{k},\mu,l,\nu,l'} h_{\mu l,\nu l'}(\bm{k}+e\bm{A}(t))\ket{\bm{k};\mu,l}\bra{\bm{k};\nu,l'}$,
where the vector potential $\bm{A}(t)=-\int^t \bm{E}_{2d}(t') dt'$
with $\bm{E}_{2d}(t)=(E_x(t),E_y(t))$.

We focus on a pulse laser
of angular frequency $\Omega$,
\begin{align}
\bm{A}(t)=\frac{E_0}{\Omega}f(t)\begin{bmatrix} \cos(\Omega t) \\ \epsilon_p \sin(\Omega t)\end{bmatrix},
\end{align}
where $f(t)$ represents a 5-cycle Gaussian envelope function~\cite{supp}
and $E_0$ approximately gives the peak electric-field amplitude.
We set the angular frequency as $\hbar\Omega=0.3\eV$
corresponding to a mid-infrared laser widely used
in experiments (see, e.g., Ref.~\cite{Ghimire2011,Yoshikawa2017}).
The parameter $\epsilon_p$ distinguishes the polarization:
$\epsilon_p=0$ means the linear polarization along the $x$ direction
and $\epsilon_p=1$ the circular polarization.

Our simulation protocol is as follows.
At the initial time $t=t_\text{ini}$ $(\ll0)$, 
we take the ground state in which
every energy eigenstate
with negative (positive) energy is occupied (unoccupied).
Since we neglect interactions between electrons,
we numerically solve the time-dependent Schr\"{o}dinger equation
for individual occupied state under $\htot(t)$.
To reduce the computational cost, we ignore the time evolution
of occupied states well below the Fermi energy ($E<E_F-5\hbar\Omega$)
since their contributions to the electric current are small.
To analyze the optical response,
we consider the electric current
\begin{align}
    \hat{\bm{J}}(t) =\frac{\partial\htot(t)}{\partial\bm{A}(t)}
    =\sum_{\bm{k}} \hat{\bm{J}}(\bm{k};t)
\end{align}
and its expectation value
$\bm{J}(t)=\sum_{\bm{k}}\bm{J}(\bm{k};t)=\sum_{\bm{k}}\braket{\hat{\bm{J}}(\bm{k};t)}_t$
at each time step.
Further technical details are described in Supplemental Material~\cite{supp}.

\begin{figure}
\center
\includegraphics[width=\columnwidth]{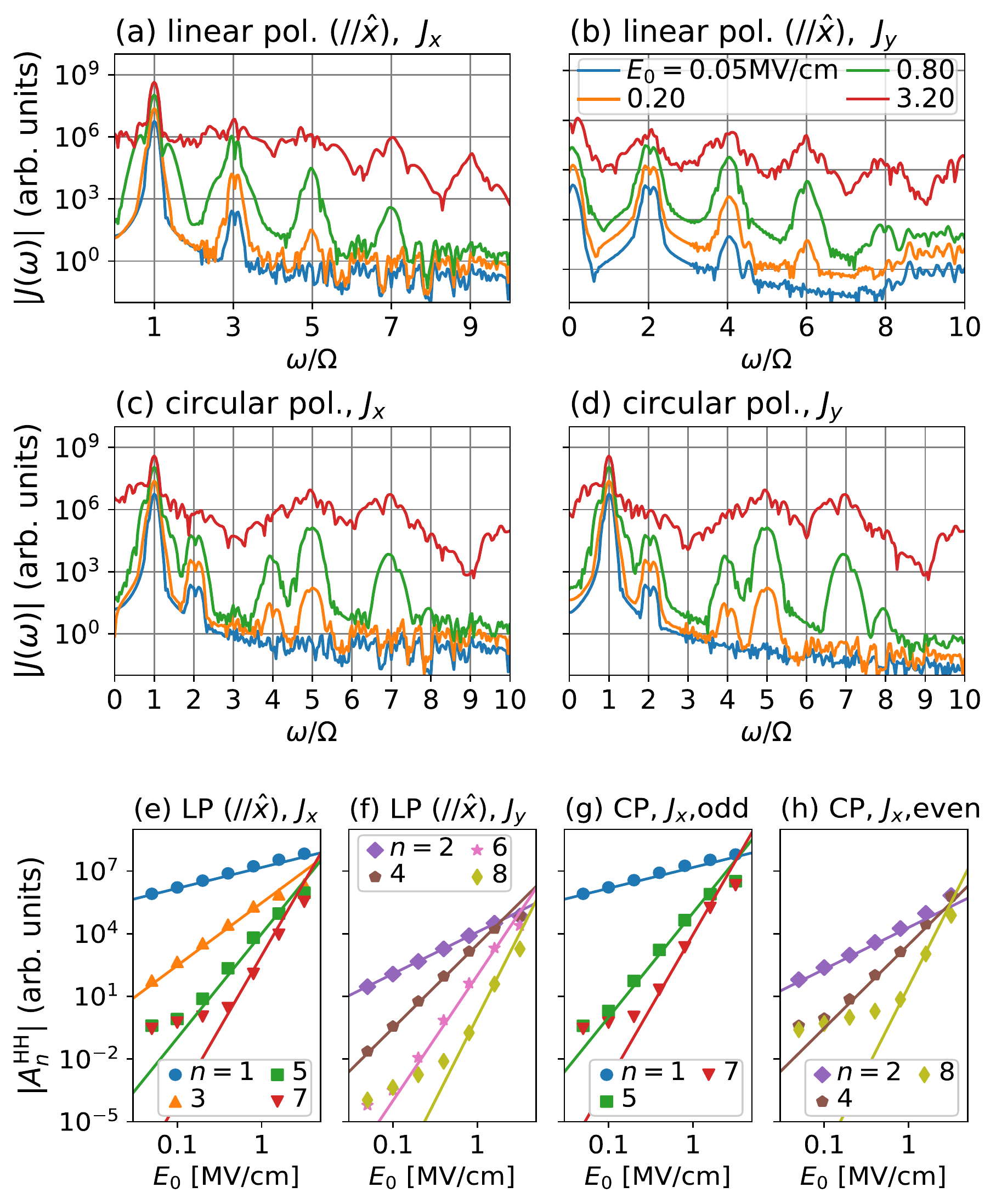}
\caption{
(a--d) Amplitude spectra for in-plane components
of electric current plotted for input electric fields
$E_0=0.05$ (blue), $0.2$ (orange), $0.8$ (green),
and $1.6\mvcm$ (red).
The polarization of the input electric field
is linear (along $x$) for (a) and (b)
and circular for (c) and (d),
and the electric-current component is $J_x$ for (a) and (c)
and $J_y$ for (b) and (d).
(e--h) Amplitudes of $n$-th harmonic $\AHH_n$ plotted against
the input field amplitude $E_0$.
In panels (e) and (h), the nonvanishing harmonics
of $J_x$ and $J_y$ for linearly-polarized fields
are plotted respectively.
In panels (g) and (h), we plot the nonvanishing
harmonics of $J_x$ in the circularly-polarized fields
at the odd and even orders respectively.
In panels (e--h), the solids lines show the eye guides
$\propto E_0^n$ for each $n$.
}
\label{fig:spectrum}
\end{figure}

{\it High-harmonic generation.}---
First, we analyze the spectra for
the electric current induced by the linearly-polarized laser.
Figures~\ref{fig:spectrum}(a) and (b) show
the spectra of the currents parallel ($\Jpara$) and perpendicular ($\Jperp$)
to the electric field, respectively.
We observe several peaks at $(2m+1)\Omega$
for $\Jpara$ and at $2m\Omega$ for $\Jperp$ ($m\in\mathbb{Z}$).
In experiments, the induced current with these harmonic peaks
is observed as the high-harmonic generation from the TBG
as illustrated in Fig.~\ref{fig:band}(a).

The even-order harmonics are characteristic
to the TBG and cannot appear in
the monolayer or conventional AA- and AB-stacked bilayers~\cite{Kumar2020}
that have inversion centers~\cite{Malard2009}
although the interlayer bias can give rise to the even-order harmonics~\cite{Brun2015,Candussio2020}.
The selection rules that $\Jpara$ ($\Jperp$)
has odd-only (even-only) harmonics
are explained by the so-called dynamical symmetry
appearing in the limit of $t_\text{FWHM}\to\infty$~\cite{Alon1998,Neufeld2019}.
Note that our TBG without the laser field
has the symmetry under $C_{2y}$, i.e., the $\pi$-rotation
about the $y$-axis (see Fig.~\ref{fig:band}(b)).
In the presence of the linearly-polarized electric field,
this symmetry is no longer true,
but $C_{2y}$ combined with the time-translation $t\to t+T/2$
becomes a symmetry transformation.
This dynamical symmetry leads to the selection rules
together with
the fact that $\Jpara$ ($\Jperp$) is odd (even)
under the transformation
(see Supplemental Material for detail~\cite{supp}).

To analyze the amplitude of the $n$-th harmonic,
we define the following quantity:
$\AHH_n\equiv\int_{(n-1/2)\Omega}^{(n+1/2)\Omega}\frac{d\omega}{\Omega}J(\omega)$,
where $J(\omega)$ represents the spectrum of some component of electric current.
In Figs.~\ref{fig:spectrum}(e) and (f), we plot the harmonic amplitude
for $n\le8$ against the incident field amplitude $E_0$.
For $E_0\le 1\mvcm$, each harmonic amplitude scales as $\AHH_n\propto E_0^n$
in line with the perturbation theory~\cite{Boydbook}.
On the other hand, in the strong-field regime $E_0\ge1\mvcm$,
$\AHH_n$ slightly saturates and deviates from the $E_0^n$-scaling.
In this regime, the harmonic peaks are not very sharp
as shown in Figs.~\ref{fig:band}(a) and (b)
due to lots of excitations occurring between the bands.

Second,
we analyze the case of the circular polarization.
The current spectra for $J_x$ and $J_y$
are shown in Figs.~\ref{fig:spectrum}(c) and (d),
in which we find a peculiar selection rule:
The harmonics at $3m\Omega$ ($m\in\mathbb{Z}$) are prohibited.
This selection rule derives from another dynamical symmetry
consisting of $C_3$, the $120^\circ$-rotation about the $z$-axis,
and the time-translation $t\to t+T/3$.
This dynamical symmetry allows the harmonics only at $(3m\pm1)\Omega$ and hence prohibits $3m\Omega$.
This symmetry argument also implies that $\bm{J}(\omega=3m\pm1)$ are circularly polarized~\cite{Neufeld2019,supp}, and thus we obtain similar harmonic peak heights for $J_x$ and $J_y$ in Figs.~\ref{fig:spectrum}(c) and (d). The harmonic amplitudes and their saturation behavior are shown in Figs.~\ref{fig:spectrum}(g) and (h).

The peculiar selection rule
under the circularly-polarized field
is characteristic of the TBG
and not present in the monolayer or conventional bilayers.
The monolayer and the AA-stacked bilayer
have the 6-fold rotational symmetry, and
thus the harmonics are allowed only for $(6m\pm1)\Omega$~\cite{Alon1998}.
The AB-stacked bilayer also allows only harmonics at $(6m\pm1)\Omega$
due to the 3-fold-rotation and inversion symmetries.
These symmetries forbid the harmonics $3m\Omega$ and $2m\Omega$, respectively,
and the allowed harmonics are only $(6m\pm1)\Omega$.
The TBG is less symmetric than the monolayer and conventional bilayers, exhibiting enriched nonlinear optical responses with orders $n=6m\pm2$.

\begin{figure}
\center
\includegraphics[width=\columnwidth]{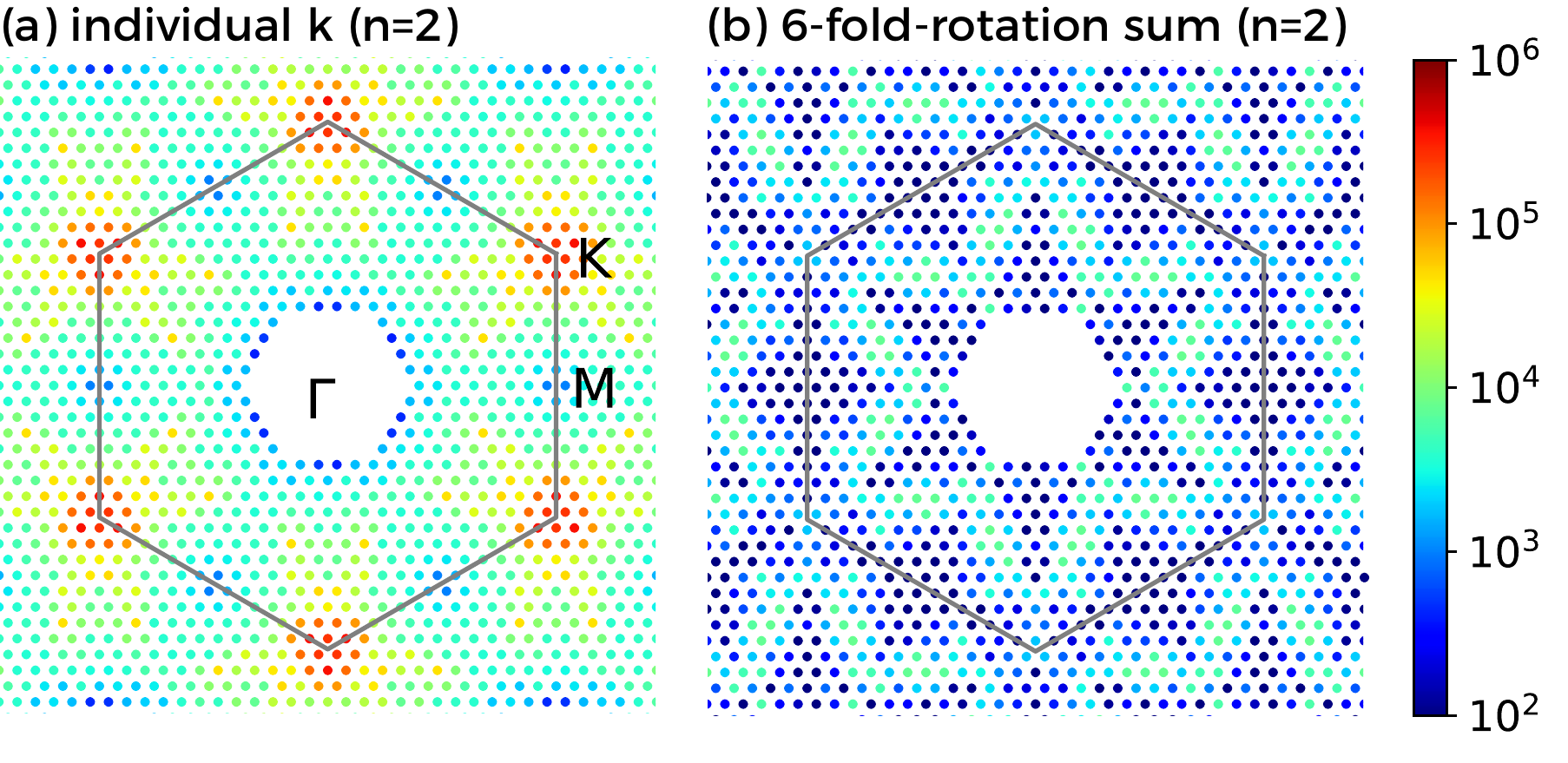}
\caption{
(a) $\bm{k}$-resolved harmonic amplitude
$|\AHH_{n}(\bm{k})|$ for $n=2$ over the $\bm{k}$-space.
(b) Absolute value of the 6-fold-rotation sum $\SHH_{n}(\bm{k})$
for $n=2$ (see text for definition).
In both panels, we use the extended zone scheme, duplicating the data outside the first BZ.
}
\label{fig:kresol}
\end{figure}

{\it Reciprocal-space analysis.}---
Having found the harmonic responses characteristic to the TBG,
we now investigate their mechanism.
To this end, we look into 
the harmonic amplitude resolved in the reciprocal space
by introducing 
$\AHH_n(\bm{k})\equiv\int_{(n-1/2)\Omega}^{(n+1/2)\Omega}\frac{d\omega}{\Omega}J(\bm{k};\omega)$,
where $J(\bm{k};\omega)$ represents some component
of the Fourier transform of $\bm{J}(\bm{k};t)$.

Figure~\ref{fig:kresol}(a)
shows the $\bm{k}$-resolved second harmonic amplitude
$|\AHH_{n=2}(\bm{k})|$ obtained for the circularly-polarized field
with $E_0=0.8\mvcm$.
The largest amplitude exists
in the vicinity of the $K$ and $K'$ points
and this tendency is commonly seen for
the other harmonic orders $n$. 
This observation means that large nonlinear currents
are carried by the Dirac electrons
(see Fig.~\ref{fig:band}(d))
consistently with the experimental
results showing that the Dirac electrons
generate harmonics very efficiently~\cite{Cheng2020,Kovalev2019}.

Nevertheless,
nonDirac electrons
play more significant roles
in the second harmonic
after the sum over the BZ.
To show this, we focus on the $6$-fold-rotation
sum of the $\bm{k}$-resolved harmonics
and define
$\SHH_{n}(\bm{k})\equiv \sum_{\ell=0}^5 \AHH_{n}(\mathcal{R}_z(\pi/3)^\ell\bm{k})$.
We note that the total harmonic amplitude $\AHH_n$
is obtained as a weighted sum of $\SHH_{n}(\bm{k})$.
Figure~\ref{fig:kresol}(b) shows $|\SHH_{n=2}(\bm{k})|$
over the $\bm{k}$-space,
in which we find that the $\bm{k}$ points
near the $K$ point give small contributions.
Indeed the individual Dirac electrons carry
large nonlinear currents,
but these currents cancel each other very strongly.
As a result, the nonDirac electrons
in the middle of the BZ
give more contributions for the second harmonic.
The importance of nonDirac electrons
are common with other harmonic orders $n=6m\pm2$
that are characteristic to the TBG,
whereas the Dirac electrons give dominant contributions
for the ordinary harmonics $n=6m\pm1$.

The band structure in Fig.~\ref{fig:band}(d) confirms this interpretation.
As noted above, the interlayer coupling, emerging as small band splittings, is more effective away from the $K$ point. Given that the interlayer coupling activates the characteristic harmonics $n=6m\pm2$, they are contributed from the $\bm{k}$ points away from the $K$ point.

\begin{figure}%[h]
\center
\includegraphics[width=\columnwidth]{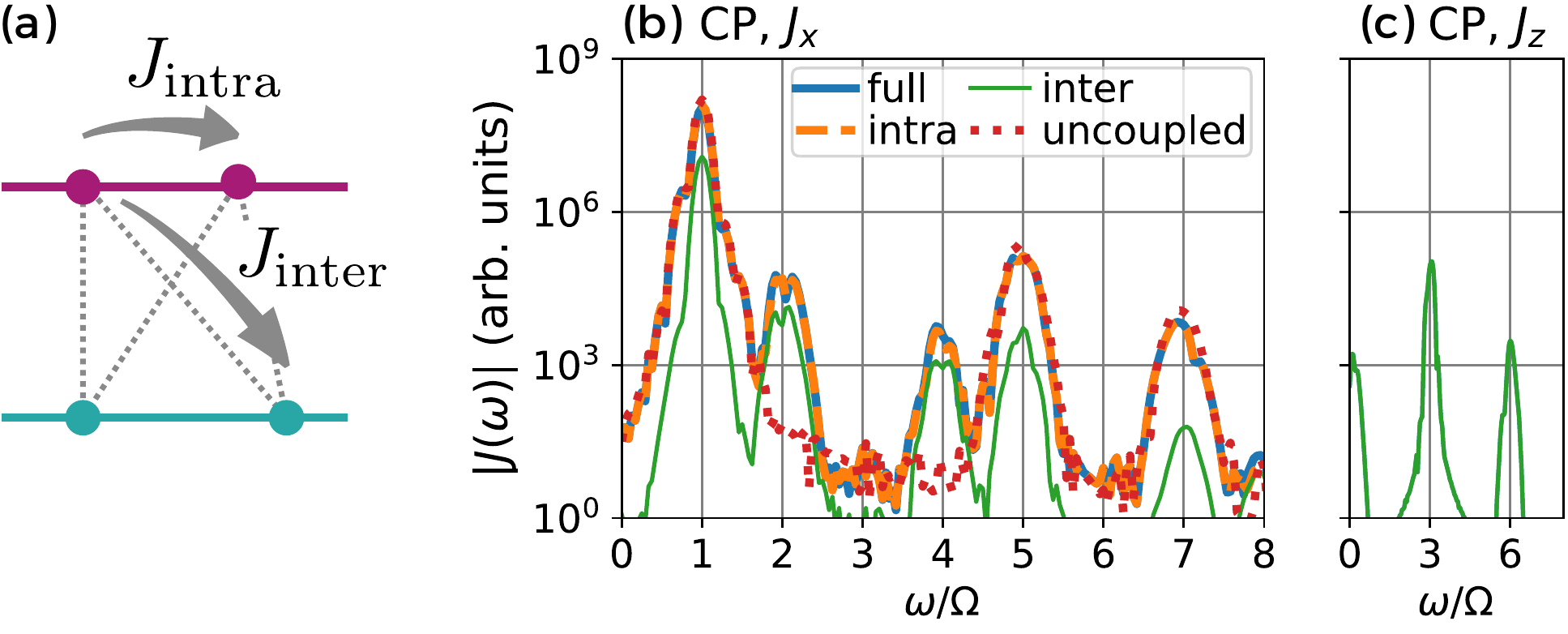}
\caption{
(a) Schematic illustration of intralayer and interlayer
electric currents in the sideview of the TBG.
(b--c) Amplitude spectra of the (b) in-plane ($J_x$)
and (b) out-of-plane ($J_z$) electric current
generated by the circular polarization with $E_0=0.8\mvcm$.
Each spectrum represents the total $\bm{J}$,
intralayer $\bm{J}_\text{intra}$,
and interlayer $\bm{J}_\text{inter}$ currents in the TBG
and the total current in the uncoupled bilayers.
}
\label{fig:inter}
\end{figure}

{\it Role of interlayer coupling.}---
To elucidate other aspects of the interlayer coupling,
we decompose the total electric current into two parts,
the intralayer and interlayer contributions,
as
\begin{align}
    \bm{J}(t) = \bm{J}_\text{intra}(t) + \bm{J}_\text{inter}(t).
\end{align}
The definitions of these contributions
follow from
the fact that the current operator $\hat{\bm{J}}(\bm{k};t)$
has a $28\times28$-matrix representation
$\hat{\bm{J}}(\bm{k};t)=\sum_{\mu,l,\nu,l'} j_{\mu l,\nu l'}(\bm{k})\ket{\bm{k};\mu,l}\bra{\bm{k};\nu,l'}$.
We define the operators $\hat{\bm{J}}_\text{intra}$ and 
$\hat{\bm{J}}_\text{inter}$ as the $l=l'$ and $l\neq l'$
parts of $\hat{\bm{J}}(\bm{k};t)$, respectively,
and 
$\bm{J}_\text{intra}(t)$ and
$\bm{J}_\text{inter}(t)$ are their expectation values.
Figure~\ref{fig:inter}(a) schematically illustrates
$\hat{\bm{J}}_\text{intra}$ and 
$\hat{\bm{J}}_\text{inter}$,
which are the electric currents accompanied by
the intralayer and interlayer hoppings of electrons, respectively.

The intralayer component
gives the dominant contribution
as shown in Fig.~\ref{fig:inter}(b),
which shows the result
for the circular polarization with $E_0=0.8\mvcm$.
Since the $x$ and $y$ components are essentially equivalent
for the circular polarization, we plot only the $x$ component.

For comparison, we also plot the result for the uncoupled bilayers
which are defined by removing all the interlayer hopping,
i.e., setting $t(\bm{R}
^{(l)}_i,\bm{R}^{(l')}_{i'})=0$ for $l\neq l'$.
Similarly to the monolayer,
the uncoupled bilayers only give the harmonics at $n=6m\pm1$.
For these harmonics, the difference between the TBG and uncoupled bilayers is quite small, meaning that they are carried by the electrons
accelerated within each layer.

Remarkably, the dominance of the intralayer current holds also for the harmonics $n=6m\pm2$ that are caused by the interlayer coupling.
Indeed the interlayer coupling is important
and, as shown in Fig.~\ref{fig:inter}(c),
there occurs significant charge transfer
between the layers
including some dc ($0\Omega$) component
corresponding to
the photogalvanic effect~\cite{Gao2020,Otteneder2020}.
However, the in-plane currents
accompanied by the interlayer hopping
give less contribution to the total current.
Rather, the in-plane currents are contributed more by
the intralayer electron hopping,
and the interlayer coupling assists them
by breaking higher symmetry of the uncoupled bilayers
and
preventing the harmonic currents
from canceling out in the BZ.

{\it Discussions and Conclusions.}---
We have conducted the nonperturbative calculations of the laser-induced electric currents in the minimal commensurate TBG, finding higher-order harmonic responses that are not present in monolayer or conventional bilayers. In contrast to the common harmonics, these new harmonics are carried more by nonDirac electrons and caused by the interplay between the intralayer and interlayer electron hoppings. The selection rules of the harmonics are qualitatively distinct and could be tested within the current optics technology. \hl{Since the point-group symmetry of the TBG is common for most twist angles, the selection rules found here should also apply to other twist angles.} The enriched harmonics in the TBG offer versatile frequency-conversion channels for future applications. 

An important future direction toward nonlinear ``Opto-twistronics'' is to unravel the dependence on the twist angle, which has been fixed to $\theta=21.79^\circ$ in this work. Qualitative results might be different for smaller angles \hl{and lower-frequency lasers} since there occur some emergent symmetries~\cite{Zou2018,Angeli2018}.
\hl{In addition, $\theta=30^\circ$ is a particularly important twist angle, at which
the TBG becomes a quasicrystal and can accommodate symmetries prohibited in ordinary crystals~\cite{Ahn2018,Yao2018,Suzuki2019}.}
Another direction is to go into the deep nonperturbative regime with even stronger fields. In this regime, one should include relaxation due to, e.g., the interband dephasing~\cite{Golde2011} and impurity scattering~\cite{Orlando2018,Chinzei2020}. We leave these open issues for future study.

\hl{{\it Note added.}---
``Opto-twistronics'' discussed here
is also called as ``twistoptics''~\cite{Yao2020}.
}

{\it Acknowledgements.}---
This work was supported by JSPS KAKENHI Grant No. JP18K13495.

%\bibliography{../../../Bibtex/TBG_HHG.bib,supp.bib}
%

%######################################
%##### Supplementary Materials ########
%######################################

\setcounter{figure}{0}
\setcounter{equation}{0}
\setcounter{section}{0}

\onecolumngrid

\begin{center}

\vspace{1.5cm}

{\large \bf Supplemental Material:
High-order nonlinear optical response of a twisted bilayer graphene}

\vspace{0.3cm}

{\large Tatsuhiko N. Ikeda} \\[2mm]
\textit{The Institute for Solid State Physics, The University of Tokyo, Kashiwa, Chiba 277-8581, Japan}\\

\end{center}

\vspace{0.6cm}

\renewcommand{\theequation}{S\arabic{equation}}
\renewcommand{\thesection}{S\arabic{section}}
\renewcommand{\thefigure}{S\arabic{figure}}

\section{Envelope function for the pulse laser}\label{app:pulse}
The explicit form of the envelope function is given by
\begin{align}
f(t)=\exp\left[-2\ln2 \left(\frac{t}{t_\text{FWHM}}\right)^2\right].    
\end{align}
Here $t_\text{FWHM}$ is
the full width at half-maximum of the intensity $\sim\bm{E}(t)^2$
rather than the field amplitude $|\bm{E}(t)|$.
We define the cycle of the laser pulse
by the ratio $t_\text{FWHM}/T$
with $T\equiv2\pi/\Omega$.
We use the 5-cycle pulse ($t_\text{FWHM}/T=5$)
to obtain all the data presented in the main text.

Note that the time-derivative of $f(t)$
gives a minor correction to $\bm{E}(t)=-\dd\bm{A}(t)/\dd t$,
since $f(t)$ varies slowly in such a multicycle pulse.
Thus $E_0$ in the main text gives the peak electric-field amplitude
approximately.

\section{Simulation details}
We take $25\times25$ $\bm{k}$-points
according to the Monkhorst-Pack method
in which the point-group symmetry of the hexagonal lattice
is respected.
The accessible number of $\bm{k}$-points is limited by
the computational time,
but we have confirmed that the qualitative features of the numerical results
do not change by varying the number.

For each $\bm{k}$-point, we diagonalize the $28\times28$ Hamiltonian matrix
$h_{\mu l,\nu l'}(\bm{k})$, obtaining the eigenstates
$\vec{\phi}_a(\bm{k})$ and the corresponding eigenenergy $E_a(\bm{k})$ for $a=1,2,\dots$, and $28$.
We let the median of all the eigenenergies $\{E_a(\bm{k})\}_{a,\bm{k}}$ be $E_M$ ($\simeq E_F$).
For each $\bm{k}$, we pick up the eigenvalues in the range of $[E_M-5\hbar\Omega,E_M]$
and define $\Lambda_{\bm{k}}=\{a\,|\, E_M-5\hbar\Omega < E_a(\bm{k}) < E_M\}$.

The evolution is numerically solved for each $(\bm{k},a)$ with $a\in\Lambda_{\bm{k}}$.
We take the initial time $t_\text{ini}=-15T=-3t_\text{FWHM}$
where $f(t_\text{ini})\simeq0$.
With the initial condition $\vec{\psi}_a(\bm{k},t=t_\text{ini})=\vec{\phi}_a(\bm{k})$,
we numerically integrate the time-dependent Schr\"{o}dinger equation
$\ii\partial \vec{\psi}_a(\bm{k},t)/\partial t = h(\bm{k}+e\bm{A}(t)) \vec{\psi}_a(\bm{k},t)$
up to the final time $t_\text{fin}=15T=3t_\text{FWHM}$.
We use the Runge-Kutta method with the time step
$\varDelta t=(t_\text{fin}-t_\text{ini})/2^{12}$.
We have confirmed that the results show almost no change
with smaller time steps.

The time profile of the electric current $\bm{J}(\bm{k};t)$
is calculated as the sum over each initial state $\bm{J}(\bm{k};t)=\sum_{a\in\Lambda_{\bm{k}}}\bm{J}_a(\bm{k};t)$.
Here, $\bm{J}_a(\bm{k};t)$ is calculated as the expectation value
of the $28\times28$ current matrix in terms of
the solution $\vec{\psi}_a(\bm{k},t)$.
The total current (density) $\bm{J}(t)$
is obtained as the average of $\bm{J}(\bm{k};t)$
over the $25\times25$ $\bm{k}$-points.

\section{Fourier Analysis}
\begin{figure}[h]
\center
\includegraphics[width=10cm]{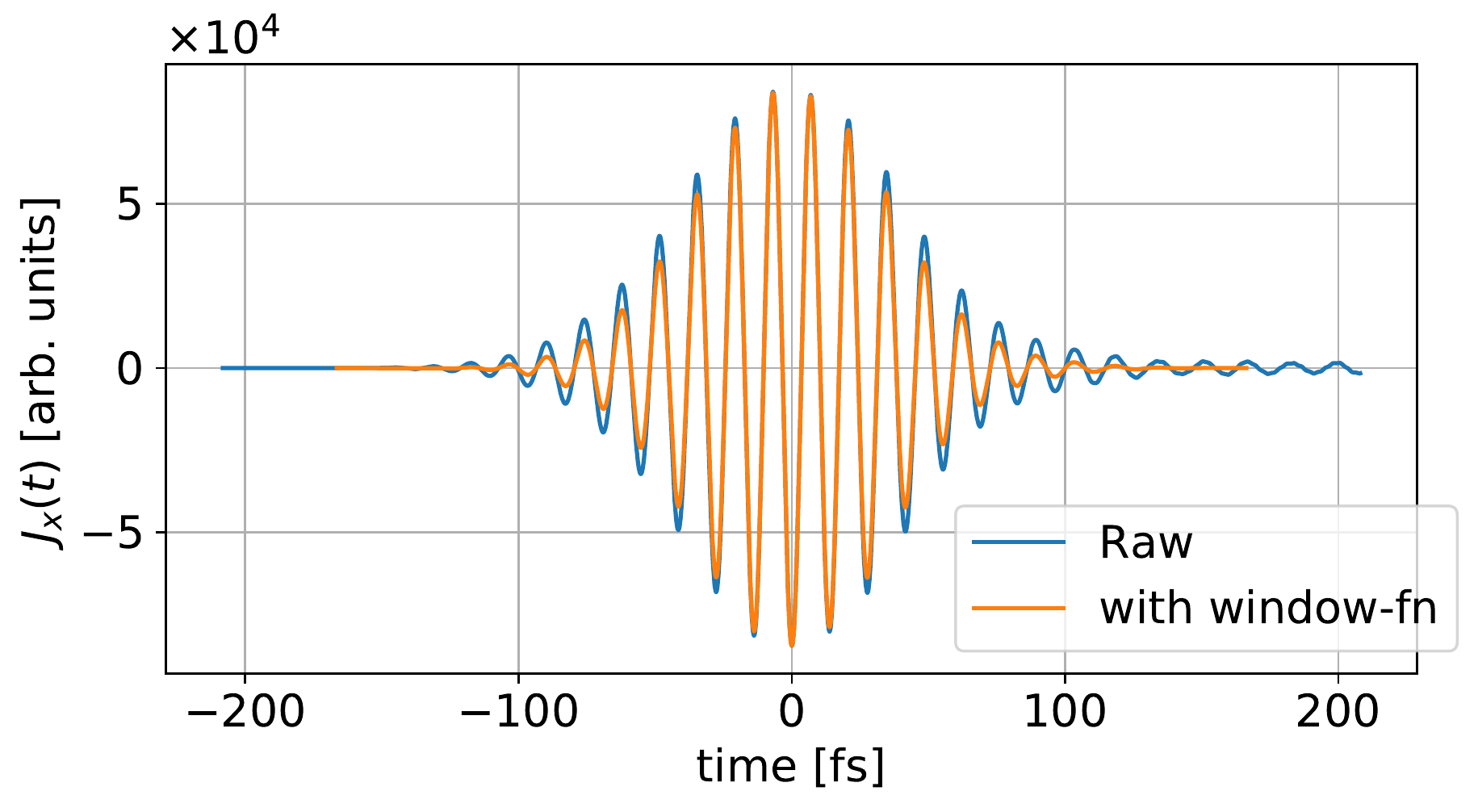}
\caption{
(Blue) Time profile of $J_x(t)$ obtained by
the simulation for the circular polarization with $E_0=0.8\,\mathrm{MV/cm}$.\\
(Orange) Modified data $\check{J}_x(t)$ with truncation and multiplication of the hanning window.
}
\label{fig:window}
\end{figure}
Since dissipation is neglected in
our calculation,
small oscillations remain after the pulse irradiation
as illustrated in Fig.~\ref{fig:window}.
This is a common technical problem
in dissipation-free models
and the standard procedure to eliminate these oscillations
is to multiply some window function to the calculated time profile
of the electric current (see e.g. Ref.~\cite{Wu2015}).%,Ikemachi2017}.

We use a similar technique
and describe the concrete procedure by taking, e.g., $J_x(t)$
in the following.
First, we discard each $10\%$ of the data
in the beginning and at the end of $J_x(t)$.
Thus we have $J_x(t)$ ($t_\text{ini}'\le t< t_\text{fin}'$)
with $t_\text{fin}'=0.8t_\text{fin}=-t_\text{ini}'$.
Second, we multiply the hanning window
\begin{align}
w(t)=\frac{1}{2}-\frac{1}{2}\cos
\left[2\pi \frac{t-t_\text{ini}'}{t_\text{fin}'-t_\text{ini}'} \right]
\end{align}
to the raw data $J_x(t)$,
and obtain $\check{J}_x(t)\equiv J_x(t)w(t)$,
which is shown in Fig.~\ref{fig:window}.
This method safely truncates the unphysical persistent oscillations
and the Fourier transform of $\check{J}_x(t)$ becomes clear
without loosing the important features in the middle of the pulse irradiation.
All the spectra in the main text are obtained by this procedure.

\section{Dynamical Symmetry and selection rule}
We prove the selection rules for harmonics based on the Floquet theory and dynamical symmetries. These selection rules and symmetries become exact when the incident field is a continuous wave, i.e., $t_\text{FWHM}\to\infty$. Thus, in the following, we assume the continuous wave and set $f(t)=1$.

\subsection{Linear Polarization}

We prove that only odd-order (even-order) harmonics
are allowed in $J_x$ ($J_y$) when the incident field is linearly polarized along the $x$ direction.
The important lattice symmetry is $C_{2y}$,
which,
at the level of Hamiltonian, means the existence
of such a $28\times28$ unitary matrix $U_{2y}$ that
\begin{align}
    h(\bm{k}) = U_{2y}h(\bm{k}')U_{2y}^\dag,~\label{app:eq:h0symm}
\end{align}
where $\bm{k}'\equiv(-k_x,k_y)$.

Now we consider the Hamiltonian in the presence of the laser field
and introduce the new notation for the Hamiltonian matrix
$h(\bm{k};t)\equiv h(\bm{k}+e\bm{A}(t))$ for clarity below.
We note that the Hamiltonian is periodic $h(\bm{k};t+T)=h(\bm{k};t)$.
Noting that $\bm{A}(t+T/2)=-\bm{A}(t)$, 
we have the following dynamical symmetry
\begin{align}
    h(\bm{k};t) = U_{2y}h(\bm{k}';t+T/2)U_{2y}^\dag.\label{app:eq:hsymm}
\end{align}
Introducing a similar notation for the current matrix by $\bm{j}(\bm{k};t)$,
we have
\begin{align}
    j_x(\bm{k};t) = -U_{2y}j_x(\bm{k}';t+T/2)U_{2y}^\dag,\label{app:eq:jxsymm}\\
    j_y(\bm{k};t) = +U_{2y}j_y(\bm{k}';t+T/2)U_{2y}^\dag.\label{app:eq:jysymm}
\end{align}

The dynamical symmetry~\eqref{app:eq:hsymm}
relates the solutions (i.e., the Floquet states)
of the time-dependent Schr\"{o}dinger equation (TDSE)
for $\bm{k}$ and $\bm{k}'$ to each other.
Let us focus on the TDSE for $\bk$:
\begin{align}
    \ii\frac{\partial \vec{\psi}(\bm{k},t)}{\partial t} &= h(\bm{k};t) \vec{\psi}(\bm{k},t).\label{app:eq:k0}
\end{align}
The Floquet theorem~\cite{Shirley1965}
dictates that the independent solutions can be written as
\begin{align}
    \vec{\psi}_a^F (\bk, t) = e^{-\ii E^F_a(\bk) t} \vec{u}_a(\bk, t),
\end{align}
where $E_a^F(\bk)$ is the so-called quasienergy
and $\vec{u}_a(\bk,t)=\vec{u}_a(\bk,t+T)$ is a periodic function
for $a=1,2\dots$, and 28.
Similarly, the TDSE for $\bk'$ is given as
\begin{align}
    \ii\frac{\partial \vec{\psi}(\bm{k}',t)}{\partial t} &= h(\bm{k}';t) \vec{\psi}(\bm{k}',t),\label{app:eq:kp}
\end{align}
and we have a set of Floquet states $\{\vec{\psi}_a^F (\bk', t)\}_{a=1}^{28}$.
Now, substituting Eq.~\eqref{app:eq:hsymm} into Eq.~\eqref{app:eq:k0}
with shifting $t\to t-T/2$,
we have
\begin{align}
    \ii\frac{\partial}{\partial t}
     [U_{2y}^\dag\vec{\psi}(\bm{k},t-T/2)]
     &= h(\bm{k}';t) [U_{2y}^\dag\vec{\psi}(\bm{k},t-T/2)].\label{app:eq:k0-2}
\end{align}
Assuming no degeneracy in quasieneries for each $\bk$
and
comparing Eqs.~\eqref{app:eq:kp} and \eqref{app:eq:k0-2},
we learn $E^F_a(\bk)=E^F_a(\bk')$
and
\begin{align}
        \vec{\psi}_a^F (\bk', t) = U_{2y}^\dag\vec{\psi}^F_a(\bm{k},t-T/2)\label{app:eq:Fconnection}
\end{align}
for each $a$ with ignoring irrelevant phase factors.
Thus the Floquet states of $\bk$ and $\bk'$
are connected to each other by the time shift and unitary transformation.

We assume that 
$\vec{\psi}_a^F (\bk, t)$ and $\vec{\psi}_a^F (\bk', t)$
are equally populated in the dynamics.
This is not exactly the case in general,
but the population imbalance typically causes
little problem (see, e.g., Ref.~\cite{Ikeda2018b}).
Then the total current consists of
the contributions from the pairwise Floquet states:
\begin{align}
    J_a^\alpha(\bm{k};t) \equiv \vec{\psi}_a^{F\dag}(\bm{k};t)j_\alpha(\bm{k};t)\vec{\psi}^F_a(\bm{k};t)
    +\vec{\psi}_a^{F\dag}(\bm{k}';t)j_\alpha(\bm{k}';t)\vec{\psi}^F_a(\bm{k}';t),\label{app:eq:Jpair}
\end{align}
where $\alpha=x$ and $y$.
As one can check easily,
Eqs.~\eqref{app:eq:jxsymm}, \eqref{app:eq:jysymm},
and \eqref{app:eq:Fconnection}
lead to
\begin{align}
    J_a^x(\bm{k};t) &= - J_a^x(\bm{k};t+T/2),\label{app:eq:Jxshift}\\
    J_a^y(\bm{k};t) &= + J_a^y(\bm{k};t+T/2).\label{app:eq:Jyshift}
\end{align}
We note, e.g., $J_a^x(\bm{k};t+T/2)=J_a^x(\bm{k};t-T/2)$
due to the periodicity.

These properties~\eqref{app:eq:Jxshift} and \eqref{app:eq:Jyshift} give the selection rules for the harmonics
as follows.
For the $x$ component, the $n$-th harmonic amplitude is
given by
\begin{align}
    J_a^x(\bk;n\Omega)
    =\int_0^T\frac{\dd t}{T}e^{\ii n\Omega t}J_a^x(\bk;t+T/2)
    =\int_0^T\frac{\dd t}{T}e^{\ii n\Omega (t+T/2)}J_a^x(\bk;t+T/2)
    =-e^{\ii n\pi} J_a^x(\bk;n\Omega).\label{app:eq:JxFourier}
\end{align}
Similarly, we have, for the $y$ component,
\begin{align}
    J_a^y(\bk;n\Omega)
    =+e^{\ii n\pi} J_a^y(\bk;n\Omega).\label{app:eq:JyFourier}
\end{align}
Equation~\eqref{app:eq:JxFourier}
means
\begin{align}
J_a^x(\bk;n\Omega)=0 \qquad (n=\text{even})    
\end{align}
whereas
Eq.~\eqref{app:eq:JyFourier} does
\begin{align}
J_a^y(\bk;n\Omega)=0 \qquad (n=\text{odd}).
\end{align}
Since we have obtained these selection rules
for each pairwise Floquet states,
we have similar rules for the total currents as well.
Thus we have proved the selection rules
for $J_x$ and $J_y$, respectively.

\subsection{Circular Polarization}
For the circularly-polarized incident field,
we have the selection rule that $3m\Omega$ ($m\in\mathbb{Z}$)
are prohibited.
To prove this, we make a parallel argument
for the linear polarization
with replacing the $C_{2y}$ symmetry to the $C_3$.
Correspondingly, the unitary matrix $U_{2y}$ is replaced
by $U_{3}$ satisfying
\begin{align}
    h(\bm{k};t) &= U_{3}h(\mathcal{R}\bm{k};t+T/3)U_{3}^\dag,\label{app:eq:c:hsymm}\\
    j_\alpha(\bm{k};t) &= \sum_{\beta=x,y}
    \mathcal{R}_{\alpha\beta}
    U_{3}j_\beta(\mathcal{R}\bm{k};t+T/3)U_{3}^\dag
    \qquad (\alpha=x,y),\label{app:eq:c:jsymm}
\end{align}
where $\mathcal{R}$ is the $2\times2$ matrix
representation of the $120^\circ$-rotation.
The pairwise Floquet states that we discussed
for the linear polarization
are now generalized to the triple-wise states
on $\bk$, $\mR\bk$, and $\mR^2\bk$.
Correspondingly, we generalize Eq.~\eqref{app:eq:Jpair}
as
\begin{align}
    J_a^\alpha(\bm{k};t) \equiv
    \vec{\psi}_a^{F\dag}(\bm{k};t)j_\alpha(\bm{k};t)\vec{\psi}^F_a(\bm{k};t)
    +\vec{\psi}_a^{F\dag}(\mR\bm{k};t)j_\alpha(\mR\bm{k};t)\vec{\psi}^F_a(\mR\bm{k};t)
    +\vec{\psi}_a^{F\dag}(\mR^2\bm{k};t)j_\alpha(\mR^2\bm{k};t)\vec{\psi}^F_a(\mR^2\bm{k};t),\label{app:eq:Jtriple}
\end{align}
which satisfies
\begin{align}
    \bm{J}_a(\bm{k};t) = \mR\bm{J}_a(\bm{k};t+T/3)
    =\mR^2\bm{J}_a(\bm{k};t+2T/3).\label{app:eq:c:Jshift}
\end{align}
These equalities lead to
\begin{align}
    \bm{J}_a(\bm{k};n\Omega)
    =\mR e^{\ii 2\pi n/3}\bm{J}_a(\bm{k};n\Omega)
    =\mR^2 e^{\ii 4\pi n/3}\bm{J}_a(\bm{k};n\Omega).\label{app:eq:c:cyc}
\end{align}
Thus, $\bm{J}_a(\bm{k};n\Omega)$ vanishes for $n=3m$
since
\begin{align}
    \bm{J}_a(\bm{k};3m\Omega)
    =\frac{1}{3}(1+\mR+\mR^2)\bm{J}_a(\bm{k};3m\Omega)=0.
\end{align}
Thus we obtain the selection rule for the circular polarization.

We remark on the nonvanishing harmonic components of $n=3m\pm1$.
From the first equality of Eq.~\eqref{app:eq:c:cyc},
we have
\begin{align}
    \mR \bm{J}_a(\bm{k};(3m\pm1)\Omega) = e^{\mp \ii2\pi/3}\bm{J}_a(\bm{k};(3m\pm1)\Omega).
\end{align}
This means that $\bm{J}_a(\bm{k};(3m\pm1)\Omega)$
are the eigenvectors of $\mR$ with eigenvalues $e^{\mp \ii2\pi/3}$.
Thus we have
\begin{align}
    \bm{J}_a(\bm{k};(3m\pm1)\Omega) \propto
    \begin{bmatrix}
        1 \\ \pm \ii
    \end{bmatrix}.
\end{align}
Namely, these harmonic currents are circularly polarized
with $\pm$ polarization.
As a consequence, we obtain
\begin{align}
    J^x_a(\bm{k};(3m\pm1)\Omega)
    = \mp \ii J^y_a(\bm{k};(3m\pm1)\Omega)
\end{align}
and similar relations
after the sums over the triple-wise Floquet states
and over the BZ.
Thus we have similar amplitude spectra
for $J_x$ and $J_y$ for the circularly-polarized
incident field.

Finally, we comment on the $z$ component
of the current.
In contrast to Eq.~\eqref{app:eq:c:jsymm},
we have
\begin{align}
j_z(\bm{k};t) &= 
    U_{3}j_z(\mathcal{R}\bm{k};t+T/3)U_{3}^\dag
    ,\label{app:eq:c:jzsymm}
\end{align}
which leads to
\begin{align}
J^z_a(\bm{k};t) = J^z_a(\bm{k};t+T/3)
\end{align}
instead of Eq.~\eqref{app:eq:c:Jshift}
and, hence,
\begin{align}
    J^z_a(\bm{k};n\Omega) = e^{\ii2\pi n/3}J^z_a(\bm{k};n\Omega).
\end{align}
This equation means that $J^z_a(\bm{k};n\Omega)$
vanishes unless $n=3m$ $(m\in\mathbb{Z})$.
This is the selection rule for the $z$ component
in the circularly-polarized field.

\end{document}